\documentclass[10pt,a4paper,twocolumn]{article}
\usepackage{f1000_styles}

\usepackage{graphicx}
\usepackage{booktabs}
\usepackage{tabularx}
\usepackage{float}
\usepackage{caption}
\usepackage{stfloats}

\DeclareGraphicsExtensions{.pdf,.png,.jpg}

\usepackage{apacite}
\usepackage[round]{natbib}
\usepackage[pdftex,colorlinks=true, linkcolor=blue, citecolor=blue, urlcolor=blue]{hyperref}
\newcommand*{\doi}[1]{\href{http://doi.org/#1}{http://doi.org/\detokenize{#1}}}

\let\cite\citep

\renewcommand{\versionDate}{\footnotesize \textit{International Journal of Human–Computer Interaction} 2024 - DRAFT ARTICLE (PRE-SUBMISSION)}

\newcommand{\authorname}[1]{\textcolor[HTML]{07608F}{\bfseries\fontfamily{phv}\selectfont #1}}  

\begin{document}
	
\title{Augmented Web Usage Mining and User Experience Optimization with CAWAL's Enriched Analytics Data}

\author[1*]{Özkan Canay}
\author[2]{Ümit Kocabıçak}
\affil[1a]{Institute of Natural Sciences, Sakarya University, Sakarya, Türkiye}
\affil[1b]{Vocational School of Sakarya, Sakarya University of Applied Sciences, Sakarya, Türkiye}
\affil[2a]{Faculty of Computer and IT Engineering, Institute of Natural Sciences, Sakarya University, Sakarya, Türkiye}
\affil[2b]{Turkish Higher Education Quality Council, Ankara, Türkiye}

\maketitle
\thispagestyle{fancy}

\begin{abstract}
	
Understanding user behavior on the web is increasingly critical for optimizing user experience (UX). This study introduces Augmented Web Usage Mining (AWUM), a novel methodology designed to enhance web usage mining and improve UX. The approach is based on enriching the interaction data provided by CAWAL (Combined Application Log and Web Analytics), a web analytics model and framework, and utilizing these datasets in web mining. In this research, more than 1.2 million session records, collected over one month by CAWAL, were processed and transformed into 8.5 GB of enriched data. These datasets were analyzed using AWUM to explore session structures, page requests, service interactions, and exit methods. The results showed that 87.16\% of sessions involved multiple pages, contributing to 98.05\% of total pageviews. Moreover, 40\% of users accessed various services, while 50\% opted for secure exits, especially when dealing with personal or sensitive information. Association rule mining revealed key patterns in frequently accessed services, offering valuable insights into service integration and user preferences. These findings reveal CAWAL’s superiority over conventional methods in precision and efficiency, providing a more comprehensive understanding of user behavior. AWUM demonstrates strong potential for advancing web mining in large-scale, multi-server environments and supports the development of more effective UX strategies through detailed interaction analysis.

\end{abstract}

\section*{Keywords}

Web usage mining (WUM); augmented WUM; user experience (UX); behavior analysis; enriched analytics data

\vspace{10pt}
\noindent\textbf{Paper type} Research paper

\vspace{30pt}
\noindent{}* Corresponding author (canay@subu.edu.tr)

\clearpage

\setlength{\parindent}{15pt} 

\section{Introduction}

Tracking visitor interactions and analyzing service usage are crucial for strategic decision-making processes in enterprise web applications and portals offering multiple services simultaneously. Web Usage Mining (WUM) analyzes user behavior and extracts meaningful patterns from this behavior to inform strategic decisions. These patterns help predict future user trends and ensure that decisions are based on solid foundations \cite{Choudhary2023}. Timely and accurate analyses provide valuable insights for developers and managers by offering concrete data to optimize web portal performance \cite{Pastorino2019}. However, understanding how users navigate between different services and interact with them in such complex systems becomes challenging with traditional methods, mainly because of large data sets and growing interaction diversity \cite{Latha2023}.

The conventional WUM process primarily relies on access logs from web servers, which typically contain basic information about which pages users access. However, these logs are insufficient for accurately defining user sessions and do not provide details on interactions within pages \cite{Jin2022,Canay2023}. The limited structure of logs complicates the data analysis process, prolongs data processing times, and reduces the accuracy of analyses \cite{Abilio2021}. These limitations hinder efforts to improve the user experience and negatively affect strategic decision-making mechanisms \cite{Husin2022}. While various methods have been proposed in the literature to overcome these limitations, these methods have not been fully effective, mainly when applied to high-volume data.

Although the development of big data technologies and cloud-based solutions can potentially increase the efficiency of WUM processes, they present significant challenges regarding data security, privacy, cost, and performance. The incompatibility of traditional server logs with cloud systems complicates the data analysis, and real-time analysis performance can suffer \cite{Mehrtak2021}. Processing large-scale datasets in cloud environments increases costs, while slower data transfer speeds and processing delays limit performance. Furthermore, in decentralized cloud environments, data security and privacy become even more critical, and storing user data in such environments increases the risk of data breaches \cite{Ageed2020,Mustafa2022}. For these reasons, WUM processes involving extensive data sets face severe technical and financial barriers.

Several significant studies in the literature focus on WUM and user behavior tracking, mainly aimed at understanding user interactions within web applications. Research on web session clustering and user navigation behavior proposes methods for analyzing user movements and integrating this information into business development strategies \cite{Bayir2022,Sharma2021}. However, the inability of traditional WUM methods to accurately define sessions in large datasets and analyze some interactions reduces data quality, adversely affecting strategic decision-making processes \cite{Husin2022}. Additionally, studies on cloud-based solutions highlight issues related to data security, privacy, and costs, further driving the search for more effective solutions \cite{Miller2022,Pang2023}. Research on how WUM techniques can contribute to UX optimization emphasizes the importance of analyzing user interactions to improve UX and suggests developing strategies accordingly \cite{Huidobro2022,Gayatri2022}.

The CAWAL framework, proposed by \citet{Canay2024} and developed as a model combining application logs with web analytics, enables more comprehensive and accurate tracking of user interactions while providing an integrated solution to current methods. Unlike traditional approaches based on web server logs, this model allows the collection and integrated analysis of more extensive data at the application level, offering higher accuracy and performance in session identification and data processing. CAWAL also provides efficiency in data processing speed and session identification processes, standing out regarding data ownership and independence, making it a sustainable and cost-effective solution for large-scale organizations.

This study aims to use the extensive session and pageview data provided by the CAWAL framework, which has been processed through sessionization and data aggregation, to generate enriched analytical data. The Augmented Web Usage Mining (AWUM) approach, developed to utilize this data as a source for the WUM process, is designed to improve data quality and enhance the efficiency of analytical methods. This approach is expected to enhance resource use efficiency, speed, and result accuracy, providing more precise insights from large datasets than conventional WUM processes.

The research proposes that the enriched analytical data provided by the CAWAL framework and the AWUM approach will offer higher accuracy and performance than traditional WUM methods. The use of the CAWAL model in the WUM process and the effectiveness of the AWUM approach will be tested based on the following hypotheses:

H1: The enriched analytical data the CAWAL Framework provides offers higher data accuracy than traditional web server logs.

H2: The AWUM approach increases process efficiency by eliminating the pre-processing phase in traditional WUM processes.

H3: The enriched data provided by AWUM allows for a more in-depth analysis of user behavior, offering higher accuracy in optimizing the user experience (UX).

The capacity of the CAWAL framework to analyze user behavior is the primary focus of this study, and how these analyses can optimize web portals to meet user needs is examined. In this context, the framework's role in data collection and analysis, as well as its impact on user experience (UX), is evaluated in detail. Additionally, the advantages of CAWAL in terms of data security, performance, and cost-effectiveness are reviewed, with specific examples illustrating how these advantages can be applied to large-scale enterprise web portals.

\vspace{2em}

The primary contributions of this study are as follows:

\begin{enumerate}
	\item  The AWUM approach was introduced, integrating enriched analytics data from the CAWAL framework for a more accurate and detailed analysis of user behavior compared to traditional methods.
	
	\item  The CAWAL framework simplifies the web usage mining process by removing the need for extensive pre-processing and delivering structured and high-quality data directly from application logs and web analytics.
	
	\item  CAWAL's enriched datasets improve the accuracy and efficiency of machine learning, user behavior prediction, and classification models.
	
	\item  The findings contribute to UX optimization by offering deep insights into user interactions, which can be used to enhance web portal performance and guide strategic decisions across various web services.
\end{enumerate}

The structure of this paper is organized as follows. Section 2 reviews the literature on web usage mining and user experience, evaluating the limitations in addressing complex user behaviors and assessing previous work in these fields. Section 3 explains the application of the CAWAL framework in web usage mining through the AWUM approach. This section also addresses data reliability, privacy concerns, and the construction of enriched datasets. Section 4 presents four distinct, user experience-centered analyses focusing on user engagement, navigation patterns, exit methods, and service transitions to identify user behaviors and improve interaction with the web portal. Section 5 discusses the results, highlighting AWUM's advantages over traditional methods and their impact on user experience optimization. Finally, Section 6 concludes the paper by reflecting on the implications of the findings and proposing directions for future research.

\section{Related work}

Web Usage Mining is a process aimed at understanding user behaviors by analyzing web access logs and deriving meaningful patterns from this behavior. While web analytics typically focuses only on data collection and visualization, WUM extends the analysis by incorporating data cleaning, transformation, and extracting knowledge-driven patterns \cite{Canay2023}. WUM processes typically begin with the processing of existing data rather than its collection, and their effectiveness largely depends on thorough data pre-processing. WUM studies in the literature generally rely on web server logs and focus on developing methodologies to enhance the quality of these logs \cite{Yau2020,Kumar2022a}.

\subsection{Web usage mining process and methods}

The success of WUM processes is directly linked to the accuracy of the data pre-processing phase. Traditional server logs provide semi-structured data requiring extensive cleaning and transformation steps to produce meaningful insights. Cleaning raw data and removing noise are critical steps that enhance the reliability of the analysis. \citet{Srivastava2023} have developed scalable data pre-processing methods, including heuristic techniques for robot detection, aiming to improve this stage significantly. \citet{Kaur2019} emphasized the importance and effectiveness of techniques designed to enhance web data quality. However, these methods tend to be time-consuming and resource-intensive, particularly when applied to large datasets. \citet{Ali2020} proposed a comprehensive framework for pre-processing data to model user behaviors, underlining the importance of transforming raw web log data into analyzable patterns. Moreover, \citet{Asadianfam2020} demonstrated that integrating case-based reasoning and clustering techniques into WUM processes enables a deeper analysis of user behaviors.

The discovery phase of pattern recognition is crucial in identifying user behavior patterns. Despite the challenges in pre-processing, traditional WUM methods have succeeded in clustering users based on behavior patterns, as demonstrated by \citet{Singh2021}, who found clustering algorithms effective in grouping users with similar behavior patterns. \citet{Ouf2023} successfully improved customer recommendation systems' accuracy by combining different mining techniques. These analyses focus on the ability to predict future user behaviors. \citet{Munk2021} emphasized that WUM data allow proactive design changes by forecasting user behaviors. \citet{Roy2022} proposed an efficient framework for web usage mining based on time and fairness constraints to analyze user behaviors and provide personalized recommendations promptly and fairly. \citet{Malik2021} enhanced classification accuracy by combining random forest algorithm results with ant colony optimization (ACO). \citet{Serin2022} investigated a fuzzy C-means-based reduced feature set association rule mining approach to predict web user behavior patterns. Similarly, \citet{Elsheweikh2023} proposed a new web recommendation model based on web usage mining techniques.

WUM methods have evolved significantly in recent years, particularly with the integration of machine learning algorithms, which have enhanced the accuracy of user behavior analysis. For instance, \citet{Asadianfam2020} employed case-based reasoning techniques to group users and analyze behaviors, while \citet{Singh2021} utilized clustering techniques for the same purpose. Both approaches have proven effective, yet challenges persist in managing the scale and complexity of modern web applications. In addition, WUM-based frameworks are widely applied in e-commerce websites to increase customer retention and satisfaction. For example, \citet{Waqas2018} demonstrated the value added to businesses through the analysis of web log files, which allowed the identification of user behavior patterns. Furthermore, \citet{Cahaya2020} explored the impact of Internet banking service quality on e-customer satisfaction and loyalty, emphasizing the importance of WUM-based applications in enhancing customer satisfaction.

\subsection{The role of web usage mining in user experience design}

User experience (UX) design aims to optimize websites based on user expectations and focuses on enhancing users' interactions with the site. Incorporating UX design principles into web platforms is essential for improving usability and engagement. User-centered design (UCD) plays a crucial role in this process and is vital for successful UX outcomes \cite{Lallemand2015}. UX optimization follows continuous improvement through A/B testing, usability testing, and feedback surveys. A/B testing compares the performance of different design alternatives, while usability testing analyzes users' interactions with the system. The literature emphasizes that these tests provide concrete feedback to improve design decisions by enhancing the effectiveness of UX \cite{Sowbhagya2023}.

\begingroup
\setlength{\extrarowheight}{4pt}  
\begin{table*}[!b]
	\setlength{\parindent}{0pt} 
	\centering
	\caption{Recent studies related to WUM and UX fields.}
	\label{tab1}
	
	\begin{tabular}{p{0.45\textwidth} p{0.305\textwidth} p{0.165\textwidth}} 
		\toprule
		\textbf{Key Contribution} & \textbf{Methods \& Techniques Used} & \textbf{Reference}  \\ \hline 
		\midrule

		Categorizes users based on navigation patterns in online directories to enhance user segmentation. & Association Rule Mining, Apriori, Fuzzy clustering & \citet{Athinarayanan2023} \\
		Reveals user requirements by discovering patterns from web log data through web usage mining. & Data Mining, Pattern Analysis, Clustering, Classification & \citet{Dubey2024} \\
		Develops a collaborative recommendation system model for improving user experience. & Clustering, Rule Extraction, Neural Network, Genetic Algorithm & \citet{Elsheweikh2023} \\
		Enhances web interfaces by increasing the predictability of user interactions using web usage mining. & Clustering, Session Reconstruction Algorithm, Bayesian Network & \citet{Jors2023} \\
		Discovers user navigation patterns through session clustering and measures user engagement. & Clustering, K-Means, K-Medoids, Bisecting K-Means & \citet{Lim2023a} \\
		Develops personalized recommendation systems based on user shopping preferences in e-commerce. & Recommendation Systems, Apriori Algorithm & \citet{Maheshkumar2021} \\
		Improves user experience on a government website by analyzing user behavior. & Association Rule Mining, Apriori, FP-Growth, Sequential Pattern Mining & \citet{Rawira2023} \\
		Understands user behavior and provides personalized web experiences through recommendation systems. & Clustering, Association Rule Mining, K-Means, Fuzzy C-Means & \citet{Serin2022} \\
		Analyzes user behavior in e-commerce websites using web usage mining for personalization. & Association Rule Mining, Apriori Algorithm & \citet{Soewito2023} \\
		Proposes an innovative method for user profile creation and updating to improve user experiences. & User Profiling, Cognitive Psychometric Memory Model & \citet{Sowbhagya2023} \\
		Presents a new recommendation system to improve website structure by shortening user navigation paths. & Recommendation Sys., Reinforcement Learning, Adaptive ranking & \citet{Ting2024} \\ 
		Identifies user interests on e-commerce sites and provides personalized recommendations. & Clustering, K-Means, DBSCAN, Hybrid Density-Based K-Means & \citet{Win2024} \\

		\bottomrule
	\end{tabular}
\end{table*}
\endgroup

Web usage mining (WUM) and user experience (UX) design complement each other, working together to improve user experiences by analyzing user behaviors. WUM provides UX designers valuable data for proactive design changes by analyzing users' online behaviors \cite{Ali2021}. For instance, WUM data on e-commerce platforms enables personalized product recommendations by analyzing users' purchasing patterns. This personalization enhances customer satisfaction while strengthening users' engagement with the platform \cite{Benali2022}. These studies demonstrate that incorporating WUM insights into UX strategies is essential for improving user engagement and satisfaction, particularly in large-scale web environments.

In addition to enhancing user satisfaction, WUM also facilitates personalized recommendations and plays an essential role in identifying user pain points and abandonment patterns. The importance of using interaction data to develop predictive models for future user behavior is frequently emphasized in the literature \cite{Wasino2023}. Such models empower UX designers to identify challenges and implement appropriate design improvements quickly. A successful example in this regard is the Cluster-N-Engage framework proposed by \citet{Lim2023b}, which utilizes WUM data to uncover significant opportunities for UX enhancement. 

Recent studies further demonstrate the critical role of WUM in enhancing UX by offering insights into user interactions and behaviors across various platforms. \autoref{tab1} provides an overview of these studies, highlighting their key contributions, methods, and techniques from the past four years, summarizing the latest advancements in WUM and UX integration.

\subsection{Challenges in existing WUM tools and UX integration}

Many existing WUM tools still rely on traditional web server logs, which often fail to capture the full range of user interactions, creating significant data gaps in UX optimization \cite{Menezes2014}. These logs typically lack details on specific activities and in-page behaviors, limiting the scope of analysis. \citet{Ouf2023} emphasizes that addressing these shortcomings requires more detailed data collection and advanced analytical methods. In response, \citet{Srivastava2022} introduced and evaluated preprocessing techniques, including robot detection, within their MapReduce-based parallel data preprocessing algorithm to improve the efficiency of web usage mining.

Despite recent progress, current research does not thoroughly address several critical WUM and UX optimization aspects. The integration of WUM with UX design often remains superficial, particularly when examining specific details of user interactions within applications \cite{Sharma2021}. Additionally, few studies have explored the analysis of dynamic in-page behaviors or how such data can be incorporated into UX optimization processes, which remains a significant gap \cite{Rawira2023}.

\begin{figure*}[b]
	\centering
	\includegraphics*[scale=0.58]{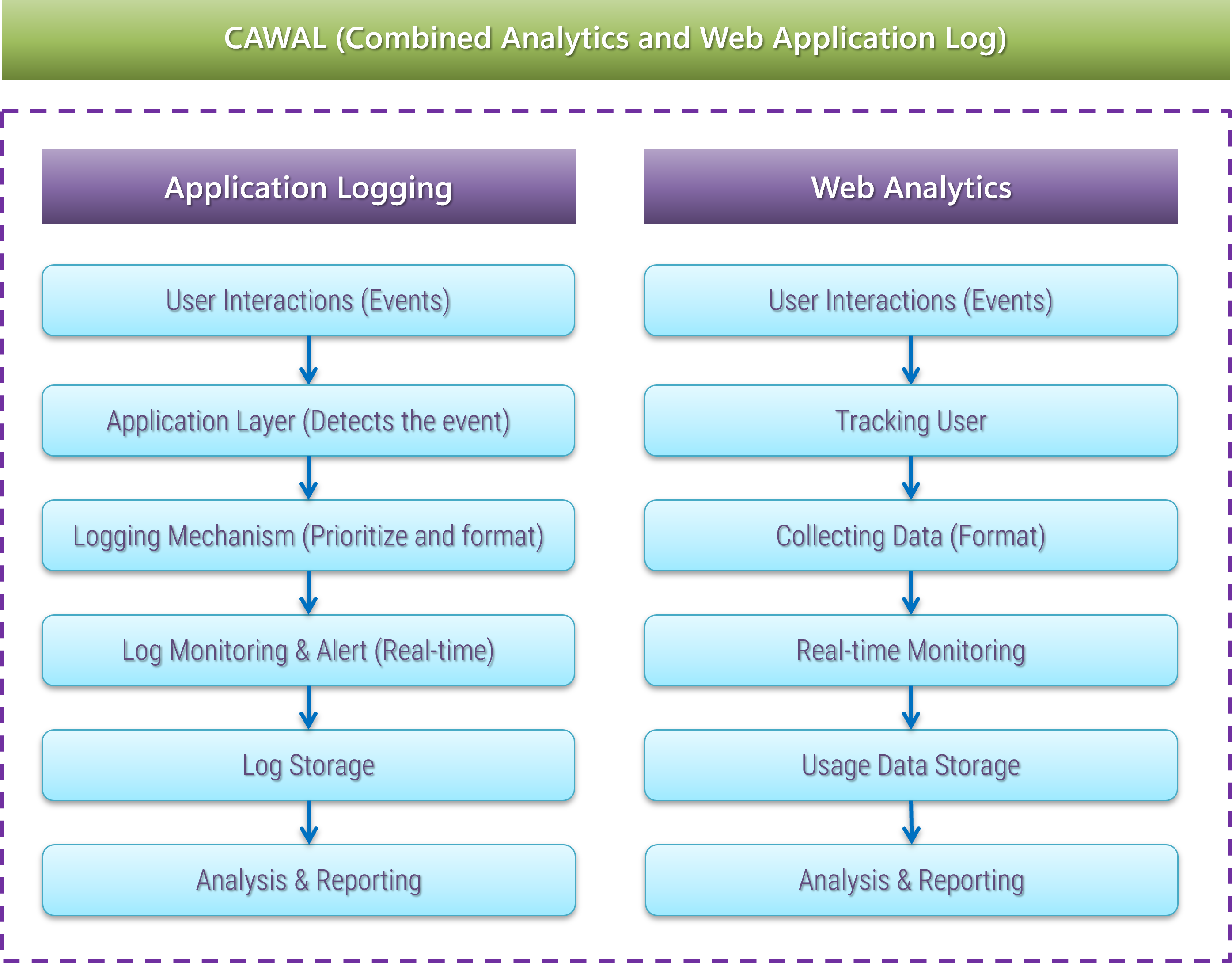}
	\caption{Application logging and web analytics processes combined by CAWAL.}
	\label{fig1}
\end{figure*}

Furthermore, there is a notable lack of research on data diversity and real-time analysis. Most studies continue to focus on static datasets, paying insufficient attention to the real-time analysis of streaming data and its implications for UX optimization \cite{Kumar2022b}. Machine learning and artificial intelligence techniques in WUM and UX integration are also underdeveloped, leaving considerable untapped potential for proactive UX improvements based on behavior prediction \cite{Xinghai2023}. Moreover, concerns about data privacy and security are becoming increasingly prominent, requiring the development of new methods to process and analyze user data securely \cite{Zagan2023}.

The CAWAL framework, central to this study, addresses many of these limitations by integrating web analytics with application-level logs and providing enriched datasets that capture complex user interactions across multiple services. This approach enhances the accuracy of user behavior tracking and facilitates real-time processing of high-dimensional data, which traditional WUM tools struggle to handle efficiently. CAWAL's sessionization and cross-service tracking capabilities offer the necessary infrastructure for detailed user behavior analysis, particularly in multi-service portals and multi-server web farms, where interactions are more complex and fragmented.

\section{Methodology}

The CAWAL (Combined Application Log and Web Analytics) model and its software framework \cite{Canay2024} implementation were integrated into CAWIS (Campus Automation Web Information System), Sakarya University's large-scale institutional web application \cite{Canay2011}. The CAWIS system, based on a portal architecture, operates within a load-balanced web farm and uses separate subdomains for each service. The interaction data collected from the CAWIS and its use in WUM processes are explained in depth, including how CAWAL eliminates the data pre-processing phase, in the following sub-sections.

\subsection{CAWAL model and framework}

The CAWAL model combines application logging with web analytics and extends the specialized data collection methodology, proposed previously \cite{Canay2023}, to a broader perspective, addressing enterprise web portals and multi-server architectures. The analytical software framework developed as a practical application of this model includes an API for data collection, a database model for data storage, and a method for generating analytical insights. This framework enables the comprehensive collection of page movement data on web portals, facilitating its use in data mining and business intelligence applications.

\begin{figure*}[b]
	\centering
	\includegraphics*[scale=0.575]{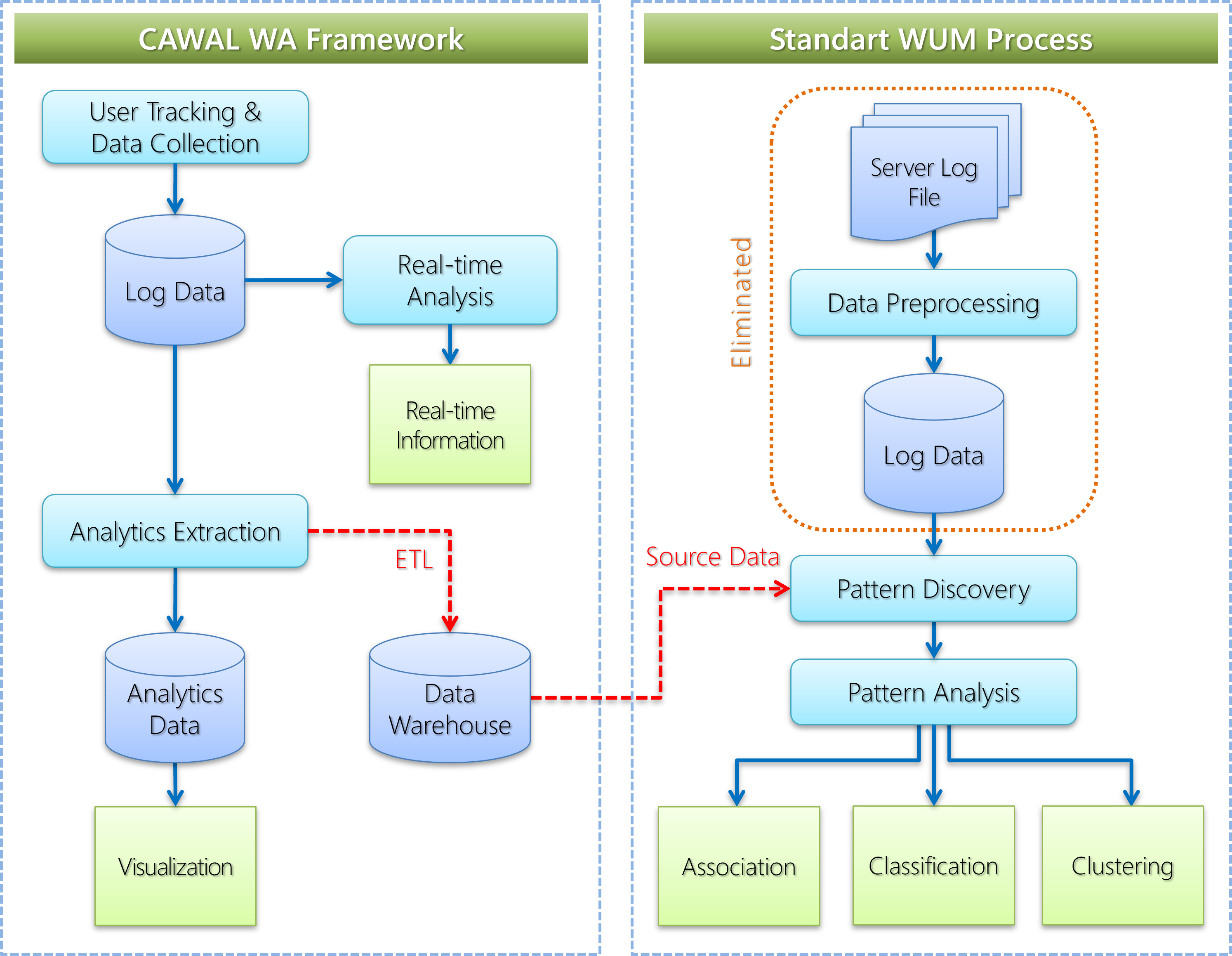}
	\caption{Usage of web access data collected with CAWAL as data source in web usage mining.}
	\label{fig2}
\end{figure*}

CAWAL was developed based on traditional software development practices and later expanded to include web analytics features. \autoref{fig1} illustrates the fundamental characteristics shared between application logging and web analytics. While conventional web analytics emerged to capture basic user interactions, modern tools now focus on gathering comprehensive data relevant to applications. In contrast, the CAWAL model was designed primarily to log application activities, with web analytics as a secondary function. This dual capability distinguishes CAWAL from other analytics tools by not only collecting detailed application logs but also integrating them into advanced web usage analytics processes.

\subsection{Integrating CAWAL into web usage mining}

WUM utilizes rich data sources to understand users' interactions with websites and to enhance user experience based on these interactions. The CAWAL framework systematically consolidates and processes these data sources, playing a crucial role in utilizing web usage data for analytical insights. The quality of the data used in mining processes is a critical factor that directly influences the accuracy and efficiency of the analysis \cite{Wang2023}. Server logs, the traditional source of information for WUM, often contain raw, semi-structured data. This type of data presents significant challenges in the pre-processing phase, which is one of the first and most critical steps in WUM \cite{Choudhary2023}.

The pre-processing stage, which includes data cleaning, transformation, and user and session identification, varies based on the project, dataset, and methods employed. It typically constitutes a substantial portion of the time and effort spent on a WUM activity \cite{Raman2021}. The pre-processing stage, carried out entirely offline, involves complex operations and requires significant effort. However, it often fails to produce precise and successful results due to data quality, technical limitations, data diversity and complexity, pre-processing methods, and human involvement \cite{Prakash2021}.

However, the session and page navigation data collected by the CAWAL model, supported with application data, are inherently accurate and well-structured. This comprehensive data set provides the necessary information for WUM and serves as a unique and high-quality data source. \autoref{fig2} illustrates how the data generated by the CAWAL model is optimized for use as a data source in WUM. This approach effectively eliminates the most critical and challenging step of the WUM process---pre-processing---while maximizing the process's success and enabling more detailed and accurate tracking of user interactions.

\begin{figure*}[b]
	\centering
	\includegraphics[scale=0.76]{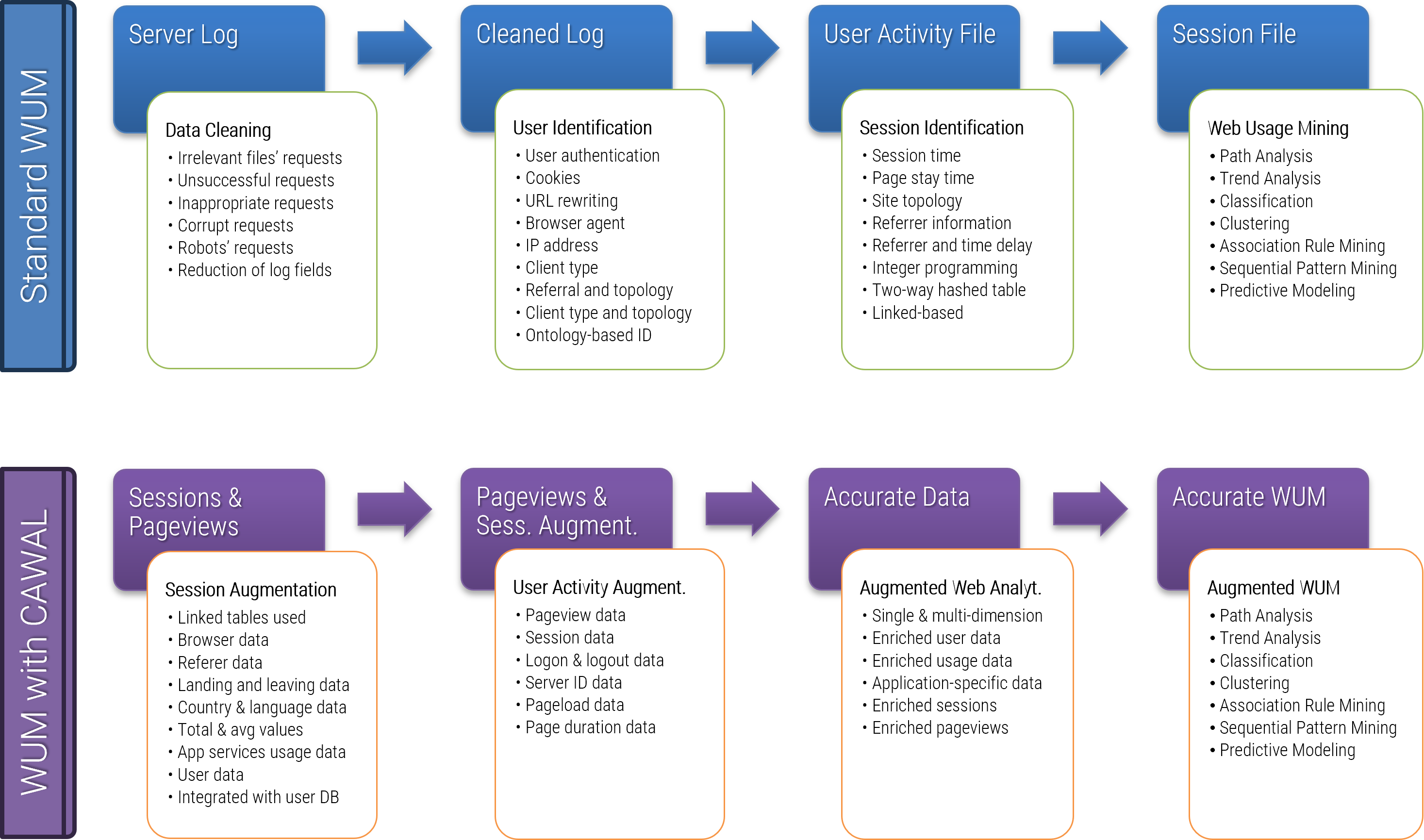}
	\caption{Standard WUM process and CAWAL's Augmented WUM.}
	\label{fig3}
\end{figure*}

\subsection{Augmented web usage mining approach}

Using high-quality, accurate raw analytical data stored in CAWAL's data warehouse eliminates the need for traditional server logs in web usage mining and renders the pre-processing step unnecessary \cite{Canay2023}. This innovative approach accelerates the analysis process by reducing the time and resources required for data cleaning, normalization, and transformation of the data collected by CAWAL, thereby improving the accuracy of the data mining results. Additionally, adopting a structured data format facilitates the understanding and the analysis of complex and challenging user behavior patterns. \autoref{fig3} compares the standard WUM process, as described by \citet{Srivastava2019}, with the WUM process implemented through the CAWAL framework.

The success of web usage mining is directly related to the quality of the raw data. In a standard WUM workflow, the procedure begins with cleaning server log data, identifying users, and determining session identities, followed by various techniques such as path analysis, trend analysis, and classification. However, the traditional method is constrained by the quality of the raw data, which may result in a loss of accuracy in the final analyses. The CAWAL framework handle these limitations through a series of data augmentation steps that enhance the richness and accuracy of the data before entering the WUM pipeline.

This process, termed Augmented Web Usage Mining (AWUM) in this study, begins with session augmentation, where connected tables, browser data, and user data are integrated to provide a more comprehensive view of user sessions and activities. It is followed by user activity augmentation, which enriches session and pageview information with detailed data such as login and logout information, server identification, and page duration. These steps culminate in enriched web analytics data, referred to as accurate data, which provides the foundation for more reliable and in-depth analyses when incorporated into the WUM process. This innovative approach enables more robust results in advanced clustering, association rule mining, and predictive modeling, allowing for a better understanding of user behavior and web usage patterns.

\subsection{Data reliability and privacy}

The absence of synthetic data in this study guarantees the reliability and integrity of the findings. All data were collected from the now-retired CAWIS web portal, developed by Sakarya University, through the CAWAL framework. The data collection process adhered to the Internet Services Usage Policy Agreement approved by the users and was fully compliant with the university's regulations. Additionally, all necessary permissions for the study were obtained from Sakarya University, ensuring compliance with both the university's internal policies and the country's legal framework. 

Diverse data anonymization techniques were applied, including the irreversible masking of user identities and the shifting of timestamps, to further safeguard user privacy. These methods effectively prevent the re-identification of user data and enhance data security. Furthermore, these adjustments aim to enhance user anonymity and reduce the risk of potential data breaches. The timestamp modification is not expected to negatively impact the analysis results, as the study focuses on trends and behavioral patterns within a specific timeframe. Every research step has been carefully conducted to safeguard participant privacy and ensure data security.

\begin{table*}[b]
	\setlength{\parindent}{0pt} 
	\centering
	\caption{Schema of the CSV file containing enriched numerical session data.}
	\label{tab2}
	
	\begin{tabular}{p{0.2\textwidth} p{0.375\textwidth} >{\raggedright\arraybackslash}p{0.155\textwidth}} 

		\toprule
		\textbf{Field Name} & \textbf{Description} & \textbf{Sample Data} \\ \hline 
		\midrule
		
		Log\_ID & Log ID number. & 12359285 \\
		Session\_ID & Session ID number. & 83665107 \\
		Log\_Date\_Time & Log date and time information. & 22.11.2022 13:00 \\
		User\_ID & User ID number. & 184922 \\
		Session\_Login\_Status & Login status in the session. & 1 \\
		Logins\_During\_Period & Number of logins during the period. & 16 \\
		User\_Type & User type (e.g. Acd./Adm./Stud./Unit). & 6 \\
		Sex & User gender (Undefined/Male/Female). & 2 \\
		Age & User age & 18 \\
		Age\_Group & User age group (Categorical, 1-4). & 1 \\
		User\_Language\_TR & Browser language. & 1 \\
		User\_Location & IP location. & 1 \\
		Browser\_Type & Web browser type. & 1 \\
		Referer\_Type & Reference type. & 6 \\
		Landing\_Srv\_ID & Service ID number first accessed. & 7 \\
		Exit\_Srv\_ID & Service ID number logged out. & 3 \\
		Exit\_Type & Logout type. & 0 \\
		Total\_Session\_Duration & Total session duration. & 730 \\
		Avg\_Page\_Duration & Average page duration. & 48.67 \\
		Total\_Page\_Load & Total page load (generation) time. & 2.88 \\
		Avg\_Page\_Load & Average page load (generation) time. & 0.19 \\
		Page\_Count & Total number of pages navigated in a session. & 15 \\
		Visitor\_PageView & Total number of pages navigated as a visitor. & 2 \\
		User\_PageView & Total number of pages navigated as a user. & 13 \\
		Service\_Count & Total number of services browsed. & 2 \\
		Page\_per\_Service & Average number of pages per service. & 7.5 \\
		Visited\_Service\_IDs & ID numbers of the visited services. & ``1,3'' \\
		\bottomrule
	\end{tabular}
\end{table*}

\subsection{Generation of enriched analytics data}

Web usage mining processes often require the handling of large and complex datasets. A comprehensive process was carried out to transform the detailed raw session and user data collected according to the data model by the CAWAL framework into enriched analytics data following sessionization procedures. Data selection and enrichment were crucial steps in preparing the data stored in the data warehouse for WUM applications. Sophisticated SQL queries created for the enriched session and pageview table views were foundational in processing data sets effectively and extracting meaningful insights.

Enriched session and pageview data were obtained through complex SQL queries performed on data marts spanning one month and totaling 8.5GB. These extensive queries, which allow for the efficient querying, filtering, and transformation of large and heterogeneous data sets, are critical for producing advanced analytics results and multi-dimensional data sets. These enriched data sets have been used to analyze user behaviors, gain a deeper understanding of user needs, and enhance the overall efficiency of the web portal's usage.

\subsection{Schemas of data files}

After the sessionization process, the enriched session data includes not only the data retrieved from the session table for each record but also summary statistics, such as the first, last, total, and average values, obtained through SQL queries. These statistics are accompanied by user-related information and pageview data corresponding to the session. The field names, descriptions, and sample data of the CSV files, which contain only numerical and categorical session information for use in web usage mining processes, are presented in \autoref{tab2}. This schema includes critical metrics such as session duration and page loads, forming the core of the data used in WUM processes.

In addition to the session data, fields summarizing user interactions with the portal services during each session are also included. Fields starting with "s\_" (e.g., s\_gate, s\_mail, s\_obis) indicate whether the user visited the respective service, with "1" representing a visit and "0" indicating no visit. Fields starting with "p\_" (e.g., p\_gate, p\_mail, p\_obis) record the number of pages the user navigated within each service. Lastly, fields beginning with "r\_" (e.g., r\_gate, r\_mail, r\_obis) represent the ratio of the number of pages visited within each service to the total number of pages visited during the session.

These fields provide detailed numerical and categorical data capturing various aspects of the sessions, which are structured for use in WUM processes. The metrics detailing session characteristics are crucial for accurately analyzing user behavior, allowing a comprehensive understanding of session structures. The enriched session data is ideal for analyzing users' overall navigation behavior throughout a session, summarizing key elements such as session duration, page loads, and visited services. This dataset plays a critical role in macro-level analyses, such as identifying patterns in session duration and service visits.

Similarly, pageview data contains detailed information about the pages visited during each user session alongside session-specific data. As a critical data source for web usage mining processes, these records include details such as the date and time of each page visit, the duration of time spent on the page, the page load time, and the page's unique identifier. Pageview data is also enriched with demographic information retrieved from the session table, including user type, gender, age group, browser type, and IP location. The reciprocal relationship between session and pageview data defines the terms used to describe these datasets, while the resulting enriched data schemas enhance the effectiveness of WUM activities conducted on it.

\subsection{Preparation of time-frame-based datasets}

As a result of the executed queries, all data related to sessions and pageviews were meticulously collected to analyze complex relationships and reveal multi-layered data structures. This data was organized through views created in the database. The data, reprocessed to include numerical and categorical information for WUM activities, was saved as separate views. In the final stage, the data corresponding to the periods specified in \autoref{tab3} was extracted from these views and transformed into enriched session and pageview datasets. These datasets, amounting to 8.5 GB collected over a one-month period, were structured and stored in CSV format to facilitate easy processing using data analysis tools such as Python.

\begin{table*}[t]
	\setlength{\parindent}{0pt} 
	\centering
	\caption{CSV data files and their properties where enriched session and pageview data are stored.}
	\label{tab3}

	\begin{tabular}{p{0.55in}p{1.5in} >{\raggedright\arraybackslash}p{0.65in}p{0.6in}p{0.5in}}
		 
		\toprule
		\textbf{Time Frame} & \textbf{Time Range} & \textbf{File Name .CSV} & \textbf{Record Count} & \textbf{File Size (MB)} \\  \hline 
		\midrule
		1-week & 2022-11-21 - 2022-11-27 & va\_page4 & 3,158,694 & 266.00 \\ \addlinespace
		1-month & 2022-11-01 - 2022-11-30 & va\_sess5 & 1,220,916 & 235.00 \\
		\bottomrule
	\end{tabular}
\end{table*}

After being collected through the CAWAL framework, the enriched session and pageview data play a pivotal role in analyzing user behavior over time. The enriched session data capture all actions performed by users during their sessions, including essential metrics such as session duration, the number of pages visited, and login-logout activities. For instance, the session data collected over one month comprises 1,220,916 records, while a one-week pageview dataset includes 3,158,694 records. The enriched pageview data provide detailed information on each visit, including metrics like pageview duration, load time, and user login status. These datasets are optimized to enable in-depth analysis of user behavior over specific time frames. Storing the data in CSV format facilitates quick and efficient processing, ensuring high accessibility for web usage mining applications.

\section{User experience-driven analysis and results}

Four analyses were conducted on enriched analytics data via the Augmented Web Usage Mining (AWUM) approach to enhance the efficiency of the web portal and optimize the user experience. These analyses were performed using Python programming language to address user engagement, experience, portal navigation, and interaction.

The first analysis examines the frequency of user exits and their correlation with specific user attributes, aiming to identify exit patterns across different user profiles. The second analysis focuses on the methods users employ to leave the system, identifying disruptions in user experience and proposing improvements to mitigate these issues. The third analysis evaluates the transition paths between portal services and their impact on user experience. The fourth and final analysis uncovers behavioral patterns by extracting association rules from user interactions.

These analyses provide detailed modeling of user behaviors, such as session transitions, pageviews, and service usage patterns, employing web usage mining techniques, including pattern analysis, user segmentation, behavioral modeling, service transition analysis, and association rule mining.

\subsection{Analysis of bounce rate and client attributes}

Bounce rate, a significant web analytics metric, represents the percentage of sessions where users view only a single page on the website and leave without any further interaction. This rate is a critical indicator for understanding users' engagement levels with the content and the impact of website design on user experience. The detailed statistical analysis of the monthly dataset reveals the differences between single-page and multi-page sessions, focusing on metrics such as the number of sessions and pageviews.

Out of the total 1,220,916 sessions in the examined dataset, 156,707 are single-page sessions, while 1,064,209 are multi-page sessions. Single-page sessions constitute 12.84\% of total sessions, while multi-page sessions make up 87.16\%. For pageviews, single-page sessions directly correspond to the session count (156,707 pageviews), while multi-page sessions generate a total of 7,882,632 pageviews. This analysis indicates that 1.95\% of the total pageviews are from single-page sessions, and 98.05\% are from multi-page sessions.

These findings provide significant insights into user engagement levels and their connection with the content on the site, offering critical data for evaluating the website's user experience. The distribution of the four categorical attributes used to distinguish users during their server access and their distribution between single-page and multi-page user groups is presented in detail in \autoref{tab4}.

\begin{table}[t]
	\setlength{\parindent}{0pt} 
	\centering
	\caption{Distribution of client attributes according to sessions with single and multiple-page visits.}
	\label{tab4}
	
	\begin{tabular}{p{1.8in} >{\raggedleft\arraybackslash}p{0.5in} >{\raggedleft\arraybackslash}p{0.5in}} \hline 
		\toprule
		\textbf{Client Attributes \newline (Categories)} & \textbf{Single Page} & \textbf{Multiple Pages} \\ \hline 
		\midrule
		Browser\_Type: &  &  \\
		\hspace{1em} 1-Standard Browser & 47.55\% & 99.17\% \\
		\hspace{1em} 2-Search Engine & 14.63\% & 0.01\% \\
		\hspace{1em} 3-Text-Based Browser & 37.82\% & 0.82\% \\
		Referer\_Type: &  &  \\
		\hspace{1em} 1-From Homepage & 1.88\% & 7.62\% \\
		\hspace{1em} 2-From Portal Services & 20.32\% & 15.44\% \\
		\hspace{1em} 3-From Corporate Domains & 4.02\% & 16.95\% \\
		\hspace{1em} 4-From Search Engine & 4.23\% & 24.44\% \\
		\hspace{1em} 5-Other Referrers & 0.31\% & 1.25\% \\
		\hspace{1em} 6-No Referrer & 69.24\% & 34.29\% \\
		User\_Language\_TR: &  &  \\
		\hspace{1em} 0-Other & 1.53\% & 2.47\% \\
		\hspace{1em} 1-Turkish & 98.47\% & 97.53\% \\
		User\_Location: &  &  \\
		\hspace{1em} 0-In T\"{u}rkiye & 21.37\% & 19.61\% \\
		\hspace{1em} 1-Internal (SAU) & 77.56\% & 79.51\% \\
		\hspace{1em} 2-Outside T\"{u}rkiye & 1.07\% & 0.87\% \\
		\bottomrule
	\end{tabular}
\end{table}

The data shows that the number of single-page sessions and the page impressions generated by these sessions are much lower than the number of multi-page sessions and page impressions. These findings indicate that multi-page sessions involve significantly more interaction and content consumption on the website. According to the analyzed data, standard browsers have a high utilization rate for multi-page viewing sessions. This observation suggests that users of standard browsers engage more with the content and interact more actively with the site. Additionally, the bounce rate is higher for visits from text-based browsers and search engines, primarily due to search engine indexing robots and automated site review tools, commonly known as web crawlers. Since these browsers do not support cookies, their actions are recorded as single-page sessions within the system. However, this group should also consider portal services that meet user needs on a single page, such as the menu service, and the immediate closure of any portal page in case of accidental opening.

The significant proportion of single-page viewing sessions without a redirector indicates that users arrive at the site by directly entering URLs or using bookmarks. Many sessions originating from search engines and involving multiple pageviews show that users accessed the site via search engines rather than using bookmarks or directly typing the site name. After finding the content they were looking for, users continued to browse the site, demonstrating the effectiveness of the SEO efforts. As expected, the web portal is predominantly used by Turkish-speaking users based on the client's browser language. This observation confirms that the site's content optimization in the local language is successful and meets the needs of local users. Additionally, the location distribution reveals a predominance of in-house and in-country users, suggesting that the site is more relevant for geographically close users.

The chi-squared ($\chi^2$) test results from the detailed analysis of single-page and multi-page session data indicate statistically significant differences in the distributions across various categories. In this test, the $\chi^2$ statistic is calculated as:
\[{\mathrm{\chiup }}^{\mathrm{2}}\mathrm{=}\sum{\frac{{\left({\mathrm{O}}_{\mathrm{i}}\mathrm{-}{\mathrm{E}}_{\mathrm{i}}\right)}^{\mathrm{2}}}{{\mathrm{E}}_{\mathrm{i}}}}\] 
where \textit{O${}_{i}$}{} represents the observed frequency, and \textit{E${}_{i}$} {}is the expected frequency for each category. The degree of freedom is determined by the formula:
\[\mathrm{Degrees\ of\ Freedom\ (DoF)}=\left(r-1\right)\left(c-1\right)\] 
where \textit{r} is the number of rows and \textit{c} is the number of columns in the contingency table. A p-value of less than 0.05 indicates a statistically significant relationship between client attributes and the bounce rate. \autoref{tab5} presents the results of this analysis, showing the impact of client characteristics on the bounce rate across four categories.

\begin{table}[b]
	\setlength{\parindent}{0pt} 
	\centering
	\caption{Effects of client attributes on bounce rate.}
	\label{tab5}
	
	\begin{tabular}{p{1.3in}r r r} \hline 
		\toprule
		\textbf{Client Attributes} & \textbf{Chi2} & \textbf{p-value} & \textbf{DoF} \\ \hline 
		\midrule
		Browser\_Type & 68.19 & 0.00 & 2 \\
		Referer\_Type & 38.72 & 0.00 & 5 \\
		User\_Language\_TR & 0.00 & 1.00 & 1 \\
		User\_Location & 0.12 & 0.94 & 2 \\
		\bottomrule
	\end{tabular}
\end{table}

The test results show a statistically significant difference between single-page and multi-page sessions' distribution of browser type and reference type. The high Chi-square values and low p-values ($\mathrm{<}$0.05) for these two attributes indicate a significant relationship with the type of session (single-page vs. multi-page), suggesting that the observed differences are unlikely to be due to random variation. On the other hand, the low Chi-square values and high p-values for the user language (Turkish) and user location attributes suggest no significant relationship with the type of session, and any slight differences that exist may be attributable to random variation. The degrees of freedom (DoF) are calculated as the number of categorical levels of the tested attribute minus one since the analysis focuses on a single variable. For instance, for Browser\_Type, which has three categories (Standard Browser, Search Engine, Text-Based Browser), the DoF is derived as 3 - 1 = 2. The DoF, alongside the Chi-square value and p-value, is crucial in determining the statistical significance of the relationship.

Bounce rate is often linked to the quality of a website's initial impression and the relevance of its content, commonly seen as an indicator of user engagement. While a high bounce rate is frequently viewed negatively, it does not always suggest a poor user experience. For example, if users quickly find what they need, such as through a service menu, their immediate exit may reflect successful navigation rather than disengagement. However, when high bounce rates are concentrated on specific pages or audience segments, it can indicate areas where the user experience or content requires improvement. Understanding these patterns is essential for identifying the key factors driving user behavior and provides a solid foundation for optimizing site design and content strategies.

\subsection{Service-based analysis of exit methods}

\begin{figure*}[b]
	\centering
	\includegraphics[width=0.90\textwidth]{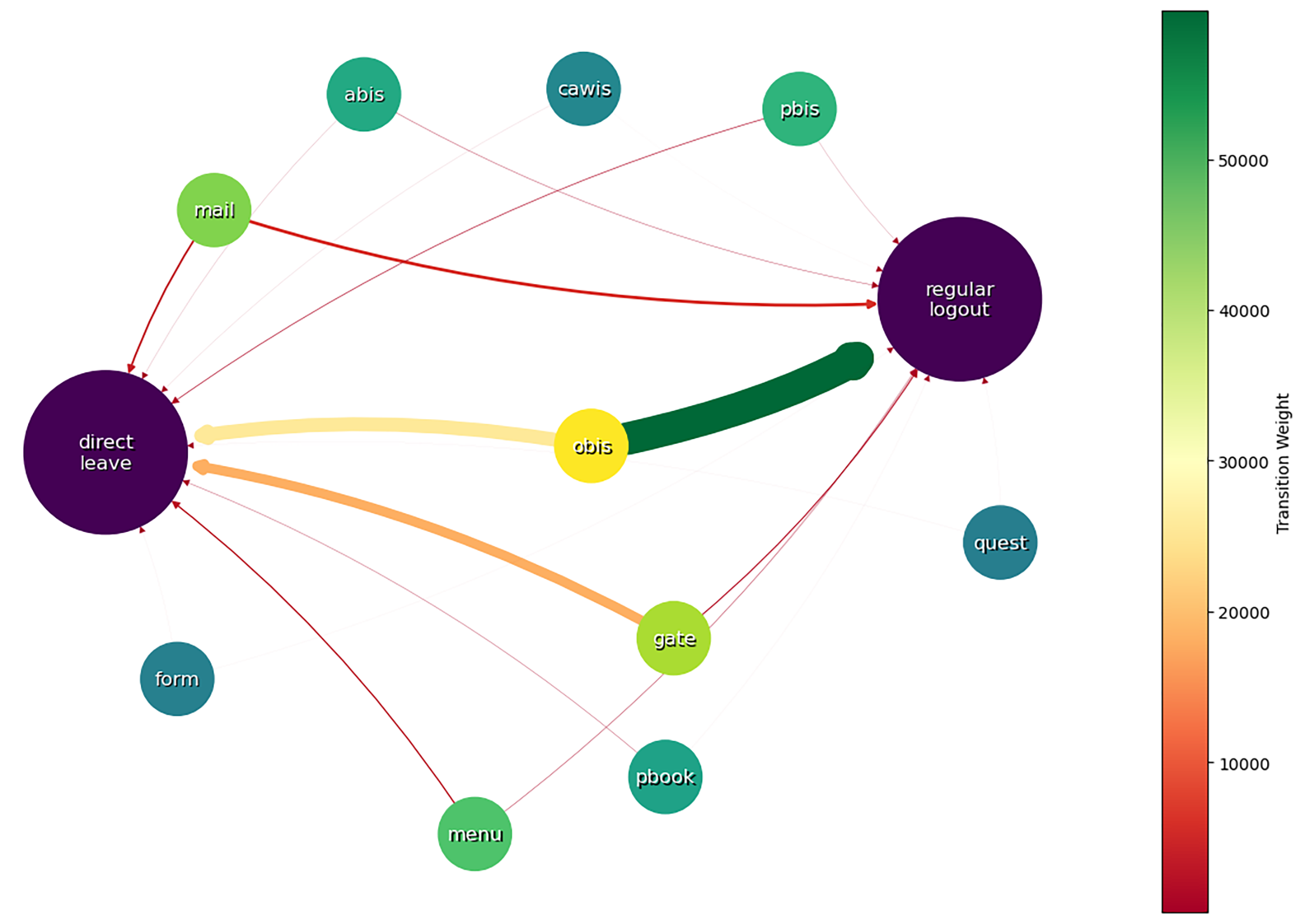}
	\caption{Service-based analysis of portal exit methods.}
	\label{fig4}
\end{figure*}

The ideal way for users to leave a web application after completing their tasks is to log out using the "secure exit" method before exiting the system. However, due to the nature of the web, it is impossible to compel users to log out securely if they choose not to. Failing to log out securely may lead to security vulnerabilities and compromise the accuracy of the collected analytical data.

The exit method analysis is mathematically modeled using a directed graph \textit{G~=~}(\textit{V},~\textit{E}), where each service is represented as a node \textit{v}~$\mathrm{\in }$~\textit{V}, and each transition between services, including exit methods, is represented as a directed edge$e=\left(v_i,v_j\right)\mathrm{\in }E$. The weight $w_{v_i\mathrm{\to }v_j}$ of each edge corresponds to the frequency of transitions from service \textit{v${}_{i}$} {}to service \textit{v${}_{j}$}, capturing user behavior and exit preferences across the web portal. This graph allows for analyzing both service transitions and preferred exit methods.

In the graph, each node's degree \textit{d}(\textit{v}) is determined by the sum of incoming and outgoing transitions, representing the activity level of each service concerning user exit strategies. The edge weights $w_{v_i\mathrm{\to }v_j}${}{}, normalized to ensure comparability, provide insights into the most common paths users take when leaving services, highlighting the relative importance of secure exits versus direct exits such as closing the browser without logging out.

\autoref{fig4}, based on the analysis of one week's session data, presents a network visualization of the methods users prefer when exiting various services on the web portal. This visualization, representing an essential data set, provides a basis for discussing user behavior trends and exit strategies on the portal and for strategic decisions to enhance the user experience.

The two primary methods for exiting the portal are "regular session termination" via the secure exit button and "direct system exit" without logging out. A small subset of session transactions, handled through the warning window, is included in the regular session termination group. Both primary methods, represented by the purple color indicating preference intensity, are used with approximately equal frequency. The color and thickness of the edges (connections) between services and exit methods represent transition weights. These transition weights are visualized by a color scale ranging from green to red, with red indicating higher frequency and greater intensity of transitions. 

When portal services examine exit preferences, using services such as "gate," "obis," "mail," and "pbis" as exit points are more pronounced, with thicker, more vivid edges indicating greater intensity. This observation indicates that users generally prefer to log out when exiting these services securely. Possible reasons for this include the critical nature of services that handle personal information, privacy concerns, and increased security awareness.

\begin{figure*}[b]
	\centering
	\includegraphics[width=0.95\textwidth]{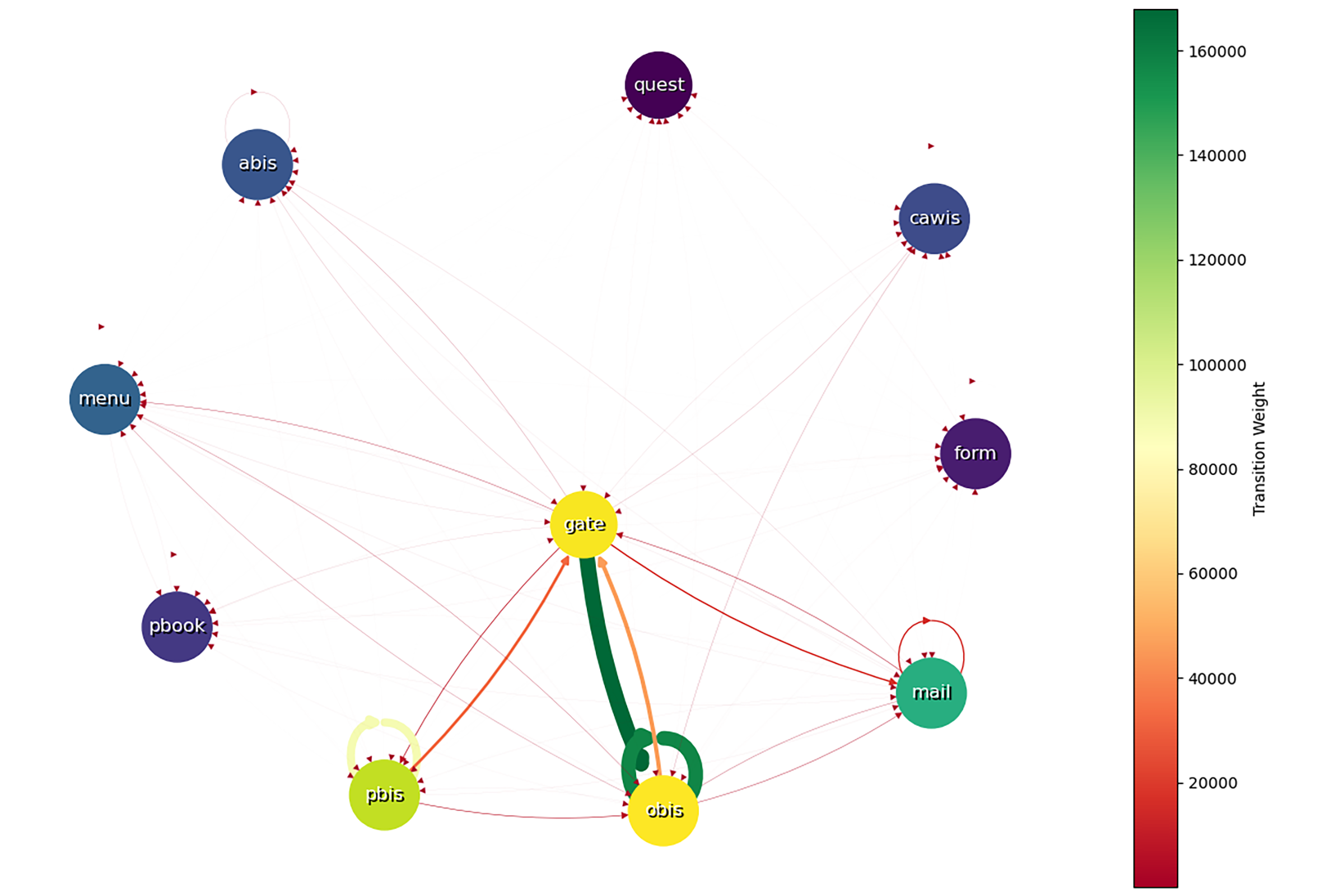}
	\caption{Transition paths and weights between portal services.}
	\label{fig5}
\end{figure*}

The analysis reveals that users mostly prefer to directly close the window or navigate to another address when exiting the "gate" service, while they are more cautious to securely log out when leaving services containing personal information, such as "obis" and "mail." The preference for secure exit in the "obis" service, which caters to students, is expected, given that some students use computers in shared institutional spaces.

Analyzing user behavior offers crucial insights for optimizing web portal design and improving user interaction. For instance, the predominant use of the secure exit option indicates that users are consciously terminating their sessions, suggesting that this function should remain easily accessible. Additionally, the significant use of the direct exit option indicates that session timeout durations or automatic logout functions should be reviewed to enhance security further.

\subsection{User interaction analysis through portal service transitions}

Analyzing user transitions between services provides valuable information about how users navigate through different sections of the web portal. Examining these interaction patterns makes it possible to identify which services are frequently accessed together and how users move between them during their sessions. Understanding these transitions is crucial for improving the overall user experience, as it can highlight potential bottlenecks, points of friction, or popular pathways users tend to follow. 

Additionally, this analysis helps uncover how effectively the portal's services are integrated and whether users encounter difficulties when switching between services. By evaluating these patterns, organizations can make informed decisions about where to focus their efforts in improving service accessibility and flow, ultimately enhancing the portal's usability. \autoref{fig5} presents a network visualization of user transitions between services in the web portal. The transitions and their corresponding weights highlight the most common navigation paths, providing a clear overview of user behavior within the portal.

The data reveal which services are most frequently utilized, as indicated by the higher transition weights between certain services. For instance, the central node "gate" is connected to many edges, indicating that it is the most frequently used service within the portal, functioning as the primary entry point into the system. In contrast, services such as "pbis," "obis," and "mail," which are also heavily used, are represented by bright nodes, while clear lines depict the transition weights to other services. The representation of bright nodes and clear lines indicates substantial user activity between these services and others.

The visualization also shows services like "cawis," "quest," and "form," characterized by darker nodes and fewer transitions. These patterns suggest that these services are utilized by a smaller subset of users, likely for specific tasks. The self-connecting arc above certain services, particularly the "gate" service, represents transitions from those services back to the system login, emphasizing the cyclical nature of user behavior in accessing and re-accessing the system.

\begin{table*}[!b]
	\setlength{\parindent}{0pt} 
	\centering
	\caption{Top 30 association rules derived using Apriori.}
	\label{tab6}
	
	\begin{tabular}{p{0.72in}p{0.48in} >{\raggedleft\arraybackslash}p{0.60in} >{\raggedleft\arraybackslash}p{0.57in} >{\raggedleft\arraybackslash}p{0.44in} >{\raggedleft\arraybackslash}p{0.58in} >{\raggedleft\arraybackslash}p{0.4in} >{\raggedleft\arraybackslash}p{0.45in} >{\raggedleft\arraybackslash}p{0.55in} >{\raggedleft\arraybackslash}p{0.40in}} 
		\toprule
		\textbf{Antecedents} & \textbf{Conse-quences} & \textbf{Antecedent\newline Support}& \textbf{Conclusion \newline Support} & \textbf{Support} & \textbf{Confidence} & \textbf{Lift} & \textbf{Leverage} & \textbf{Conviction} & \textbf{Zhangs \newline Metric} \\ \hline 
		\midrule
		abis & gate & 0.456 & 0.961 & 0.454 & 0.994 & 1.034 & 0.015 & 6.843 & 0.061 \\
		cawis & gate & 0.405 & 0.961 & 0.405 & 1.000 & 1.040 & 0.016 & infinite & 0.065 \\
		form & gate & 0.376 & 0.961 & 0.376 & 1.000 & 1.040 & 0.014 & infinite & 0.062 \\
		form & mail & 0.376 & 0.843 & 0.374 & 0.993 & 1.178 & 0.057 & 22.954 & 0.243 \\
		menu & gate & 0.474 & 0.961 & 0.474 & 1.000 & 1.040 & 0.018 & infinite & 0.074 \\
		pbook & gate & 0.433 & 0.961 & 0.433 & 1.000 & 1.040 & 0.017 & infinite & 0.068 \\
		quest & gate & 0.322 & 0.961 & 0.322 & 1.000 & 1.040 & 0.013 & infinite & 0.057 \\
		menu & mail & 0.474 & 0.843 & 0.467 & 0.984 & 1.167 & 0.067 & 9.643 & 0.273 \\
		pbook & mail & 0.433 & 0.843 & 0.430 & 0.994 & 1.180 & 0.066 & 26.412 & 0.268 \\
		quest & mail & 0.322 & 0.843 & 0.322 & 1.000 & 1.187 & 0.051 & infinite & 0.232 \\
		cawis, abis & gate & 0.302 & 0.961 & 0.302 & 1.000 & 1.040 & 0.012 & infinite & 0.055 \\
		cawis, abis & mail & 0.302 & 0.843 & 0.299 & 0.992 & 1.176 & 0.045 & 18.394 & 0.215 \\
		form, abis & gate & 0.276 & 0.961 & 0.276 & 1.000 & 1.040 & 0.011 & infinite & 0.053 \\
		form, abis & mail & 0.276 & 0.843 & 0.276 & 1.000 & 1.187 & 0.043 & infinite & 0.217 \\
		abis, gate & mail & 0.454 & 0.843 & 0.446 & 0.983 & 1.166 & 0.064 & 9.223 & 0.261 \\
		mail, abis & gate & 0.446 & 0.961 & 0.446 & 1.000 & 1.040 & 0.017 & infinite & 0.070 \\
		abis, menu & gate & 0.330 & 0.961 & 0.330 & 1.000 & 1.040 & 0.013 & infinite & 0.058 \\
		obis, abis & gate & 0.322 & 0.961 & 0.322 & 1.000 & 1.040 & 0.013 & infinite & 0.057 \\
		pbis, abis & gate & 0.400 & 0.961 & 0.397 & 0.994 & 1.034 & 0.013 & 5.992 & 0.054 \\
		pbook, abis & gate & 0.327 & 0.961 & 0.327 & 1.000 & 1.040 & 0.013 & infinite & 0.058 \\
		abis, menu & mail & 0.330 & 0.843 & 0.330 & 1.000 & 1.187 & 0.052 & infinite & 0.235 \\
		obis, abis & mail & 0.322 & 0.843 & 0.320 & 0.992 & 1.177 & 0.048 & 19.652 & 0.222 \\
		pbis, abis & mail & 0.400 & 0.843 & 0.394 & 0.987 & 1.171 & 0.058 & 12.184 & 0.244 \\
		pbook, abis & mail & 0.327 & 0.843 & 0.327 & 1.000 & 1.187 & 0.052 & infinite & 0.234 \\
		form, cawis & gate & 0.281 & 0.961 & 0.281 & 1.000 & 1.040 & 0.011 & infinite & 0.054 \\
		form, cawis & mail & 0.281 & 0.843 & 0.281 & 1.000 & 1.187 & 0.044 & infinite & 0.219 \\
		form, cawis & menu & 0.281 & 0.474 & 0.276 & 0.982 & 2.070 & 0.143 & 28.655 & 0.719 \\
		mail, cawis & gate & 0.387 & 0.961 & 0.387 & 1.000 & 1.040 & 0.015 & infinite & 0.063 \\
		cawis, menu & gate & 0.340 & 0.961 & 0.340 & 1.000 & 1.040 & 0.013 & infinite & 0.059 \\
		obis, cawis & gate & 0.343 & 0.961 & 0.343 & 1.000 & 1.040 & 0.013 & infinite & 0.059 \\

		\bottomrule
	\end{tabular}
\end{table*}

\subsection{Association rule mining on service interaction data}

Association rule mining conducted on service interaction data aims to identify patterns in user behavior and interactions to uncover services frequently used together and understand how these relationships influence the overall user experience. This analysis provides a solid foundation for designing both software interfaces and service content by revealing co-used services and their impact on various user segments.

The Apriori algorithm, one of the most commonly used methods in association rule mining, was employed in this analysis. Apriori works by iteratively identifying frequent itemsets, where the support of an itemset is the proportion of transactions that contain the itemset. The support for an itemset \textit{A} is calculated as:
\[\mathrm{Support}\left(A\right)=\frac{\mathrm{Number\ of\ transactions\ containing\ }A}{\mathrm{Total\ number\ of\ transactions}}{}\] 
After identifying frequent itemsets, association rules are generated based on these sets. An association rule is in the form of \textit{A}~$\mathrm{\to}$~\textit{B}, where \textit{A} and \textit{B} are itemsets. The confidence of an association rule is the conditional probability that a transaction containing \textit{A} also contains \textit{B}, defined as:
\[\mathrm{Confidence}\left(A\mathrm{\to }B\right)=\frac{\mathrm{Support}\left(A\mathrm{\cup }B\right)}{\mathrm{Support}\left(A\right)}\] 
Lift is another crucial metric used to evaluate the strength of an association rule, indicating how much more likely \textit{B} is to occur given \textit{A} compared to its general occurrence. It is defined as:
\[\mathrm{Lift}\left(A\mathrm{\to }B\right)=\frac{\mathrm{Confidence}\left(A\mathrm{\to }B\right)}{\mathrm{Support}\left(B\right)}\] 
A lift value greater than 1 indicates a strong positive association between \textit{A} and \textit{B}, while a value less than 1 suggests that \textit{A} and \textit{B} are negatively correlated.

The analysis was conducted using the "va\_page4.csv" dataset, which contains detailed records of user interactions with various services on the web portal. The "Service\_ID" column was first mapped to meaningful English names for easier identification of each service, as proper labeling enhances the interpretability of the results. Attributes such as user type, age group, location, browser type, and reference type were selected based on their potential influence on user interactions. These features facilitate a more precise examination of user navigation patterns and how diverse user segments interact with the portal. The service access order within each session was also determined to capture sequential patterns in user behavior, a critical step in understanding how users move through the portal's services.

The dataset was converted into transaction lists using the Transaction Encoder to facilitate the analysis. One-hot encoding was applied to prepare the data for machine learning algorithms by converting categorical data into binary format. This step ensures that the data is suitable for processing by the Apriori algorithm, which requires a binary matrix representation. The Apriori algorithm was used to identify frequent itemsets with a minimum support value of 25\% and a confidence threshold of 98\%. \autoref{tab6} presents the top thirty association rules, with lift values greater than or equal to 1, ensuring only the most significant and reliable rules are included.

The association rules outlined here reveal strong relationships between preceding services, such as "abis," "cawis," and "form," and subsequent services, such as "gate" and "mail." Metrics such as antecedent support, consequent support, support, confidence, leverage, lift, conviction, and Zhang's metric were analyzed in detail to understand users' tendencies to switch between services. For instance, the analysis found that most users who visited the "abis" service then navigated to the "gate" service, with a confidence level of 99.44\%. The leverage value of 1.0343 indicates that this transition occurs 3.43\% more frequently than would be expected by random chance. Similarly, the relationship between "form" and "mail" exhibited a notable connection, with a leverage value of 1.1784, indicating a higher-than-expected association.

The high conviction values of these rules suggest that the likelihood of the consequent occurring without the antecedent is extremely low. This finding is particularly highlighted by the infinite conviction values for transitions from "form" and "cawis" to "gate," indicating solid associations. Zhang's metric values were also evaluated to determine the reliability and direction of these relationships, confirming the robustness of these findings.

\section{Discussion}

Existing approaches in web usage mining (WUM) and user experience (UX) optimization face significant data accuracy and process efficiency limitations. The CAWAL framework employed in this study and the proposed AWUM method aim to overcome these limitations by providing an innovative solution for a deeper examination of user activity. The hypotheses of the study are based on the premise that the datasets provided by the CAWAL would offer more accurate and thorough assessments compared to conventional techniques and that these evaluations would provide strategic insights for optimizing UX.

Four critical analyses were conducted to test the hypotheses, specifically targeting different aspects of the web portal's engagements and user experience. The first analysis focused on testing the first hypothesis. The accuracy and analytical depth of the datasets provided by CAWAL were examined in detail based on users' session durations and page transitions on the web portal. The findings indicate that the framework provides more thorough and precise assessments than traditional methods. For example, out of 1,220,916 sessions, 156,707 of these, representing 12.84\%, were single-page visits, often referred to as bounce visits, while multi-page sessions made up 98.05\% of the total pageviews, demonstrating the framework's ability to analyze user activity more accurately. These results strongly support the validity of the first hypothesis.

The second and third analyses were conducted to test the second hypothesis, which assessed the CAWAL framework's potential to optimize UX. Similar to the efforts of \citet{Malik2021} to enhance random forest algorithms with ant colony optimization (ACO), this study applied a method that improves the accuracy of predictive models by enhancing data quality. In the service-based exit method examination, it was clearly observed that 50\% of users preferred secure exit methods, while the remaining 50\% predominantly used direct exit methods, such as closing the browser tab or window, navigating to another URL, or leaving the portal without logging out. The secure exit rate for services containing personal information, such as "obis" and "mail", rising to 75\% demonstrates that secure exit options are already prioritized for these services. The investigation of transitions between portal services revealed that 85\% of users used the "gate" service as the first entry point, with remarkable transitions occurring between other services. These findings clearly support the second hypothesis by illustrating the framework’s ability to capture detailed user behavior across multiple services.

One of the major contributions of this study to the literature is the introduction of the Augmented Web Usage Mining (AWUM) approach. AWUM, developed using the enriched analytical datasets provided by the CAWAL framework, offers a more extensive examination of user activity than conventional web log analysis. AWUM allows for a deeper understanding of the underlying dynamics of these engagements, moving beyond superficial analyses of user interactions with specific pages and services. These findings directly support the third hypothesis by demonstrating how enriched datasets facilitate a more detailed analysis of user behavior. For instance, user navigation patterns and interactions between frequently accessed services were examined, revealing significant insights into how specific service transitions, such as between "mail" and "obis," resulted in higher task completion rates. These insights show that AWUM’s enriched data contributes to optimizing the user experience, validating the third hypothesis.

It is essential to note that the analyses in this study contribute to the validation of all three hypotheses in a comprehensive manner. For example, the enhanced data accuracy achieved through the CAWAL framework not only improves the precision of user behavior analysis (H3) but also streamlines the entire WUM process, making it more efficient (H2). The removal of the preprocessing phase through AWUM accelerates the data analysis workflow, reducing computational complexity, which contributes to both process efficiency (H2) and improved data accuracy (H1). Additionally, the enriched datasets provided by AWUM offer valuable insights for optimizing the user experience (H3) by revealing detailed patterns in user behavior and service transitions, allowing for more informed decisions in UX design. These combined improvements highlight the overall effectiveness of the methodologies used in this study and their potential to enhance multiple aspects of web usage mining.

This effectiveness is further demonstrated by the CAWAL framework’s substantial advancements in data accuracy and processing efficiency when compared to other approaches. For instance, while the web recommendation model proposed by \citet{Elsheweikh2023} focuses on surface-level analyses of user activity, CAWAL's enriched datasets enable a more detailed and in-depth understanding across multiple services. This advanced analytical capability provides deeper insights into user engagements, making a notable contribution to the existing body of literature. Additionally, integrating the fuzzy C-means-based association rule mining method proposed by \citet{Serin2022} with CAWAL could further optimize WUM processes, facilitating more informed decision-making for UX improvement and increased user satisfaction.

Furthermore, frameworks like the time and fairness-constrained WUM approach developed by \citet{Roy2022} have made progress in developing personalized recommendation systems. However, CAWAL goes further by offering more extensive datasets and analytical capabilities, as demonstrated by the high accuracy and F1 scores achieved in predictive models. This advancement reinforces the reliability of CAWAL's assessments over traditional methods. Nonetheless, the dataset used in this study covers a specific period and user group, which may limit the generalizability of the findings.

Given these considerations, it is clear that the comprehensive data provided by the CAWAL framework enhances the precision and depth of user behavior analysis. The Augmented Web Usage Mining (AWUM) approach, developed based on this framework and supported by enriched analytical data, enables detailed examination not only of superficial user interactions but also of the underlying dynamics. This method demonstrates superior accuracy and analytical depth compared to existing approaches, positioning CAWAL as a valuable model for both academic research in the fields of WUM and UX, as well as practical applications in industries such as e-commerce, education, healthcare, and public services, where effective analysis of complex user behaviors is crucial.

\section{Conclusion and future work}

This study demonstrates that the CAWAL model significantly improves data accuracy and process efficiency in web usage mining, providing a clearer understanding of user behavior and enhancing user experience. By eliminating the time-consuming preprocessing phase, CAWAL speeds up the WUM process and reduces manual intervention. Analyses of over 1.2 million session records show that the framework offers a more precise and detailed representation of user activity, outperforming traditional methods in data accuracy. These improvements have directly enhanced the quality of user behavior analysis and service transitions.

A vital advantage of the CAWAL model is its ability to deliver practical findings through the Augmented Web Usage Mining (AWUM) approach. The enriched datasets allow for a thorough analysis of critical user behavior patterns, such as exit methods and service transitions. For instance, results show that 85\% of users started their sessions at the portal gateway, while services handling personal data, such as "obis" and "mail," had a secure exit rate of 75\%. These findings demonstrate the framework's capability to guide improvements in service integration and security features. Additionally, CAWAL ensures strong privacy protection through advanced data anonymization techniques, making it suitable for environments prioritizing data confidentiality.

Despite the promising outcomes, certain limitations must be addressed. This study relied on data from a specific period and user group, which may limit the generalizability of the findings. Future research should apply the CAWAL model to more extensive and more diverse datasets across different industries to assess its performance in various operational settings. Furthermore, exploring ways to improve data security, such as advanced encryption and secure data management, will be crucial for broader adoption in sensitive applications. Addressing these aspects will help further develop CAWAL and establish it as a valuable model for web usage mining and user experience optimization.

\setlength{\parindent}{0pt} 

\subsection*{Ethics Statement}
All data used in this study were collected from the CAWIS web portal in compliance with Sakarya University's regulations and the legal framework of the Republic of Turkey. Necessary permissions were obtained from Sakarya University, and diverse data anonymization techniques were applied to ensure user privacy and data security throughout the research process.

\subsection*{Consent Statement}
Consent for data usage was obtained through the Internet Services Usage Policy Agreement, which was approved by all users of the CAWIS web portal.

\subsection*{Disclosure statement}
No potential conflict of interest was reported by the author(s).

\subsection*{ORCID}
Özkan Canay https://orcid.org/0000-0001-7539-6001
Umit Kocabıçak https://orcid.org/0000-0003-0369-9737

\subsection*{Data availability statement }
The data are not publicly available due to privacy reasons and confidentiality agreement restrictions.

\bibliography{ref}

\bigskip

\subsection*{About the authors}

\medskip
\authorname{Özkan Canay} completed his PhD in the Department of Computer Engineering at Sakarya University, Türkiye, under the supervision of Prof. Dr. Ümit Kocabıçak. His research focuses on information systems, web usage mining, data analytics, user experience (UX) design and machine learning.

\medskip
\authorname{Ümit Koçabıçak} served as a faculty member in the Department of Computer Engineering at Sakarya University, Türkiye, for many years. He is currently the president of the Turkish Higher Education Quality Council. His research interests include data analytics, machine learning, image processing, and computational modeling in engineering.






\end{document}